\documentclass[12pt]{article}

\usepackage{graphics,epsfig,citesort,amssymb}
\usepackage{geometry}
\newcommand{\be}{\begin{equation}}
\newcommand{\ee}{\end{equation}}
\newcommand{\beq}{\begin{eqnarray}}
\newcommand{\eeq}{\end{eqnarray}}

\newcommand{\mc}{\mathcal}

\newcommand{\nn}{\nonumber}
\newcommand{\gev}{\mathrm{~GeV}}
\newcommand{\XLF}{\mathrm{\chi LF}}

\def\simge{\mathrel{\rlap{\raise 0.511ex \hbox{$>$}}{\lower 0.511ex
\hbox{$\sim$}}}}
\def\simle{\mathrel{\rlap{\raise 0.511ex \hbox{$<$}}{\lower 0.511ex
\hbox{$\sim$}}}}

\begin{document}
\thispagestyle{empty}
\begin{flushright}
\begin{tabular}{l}
{\tt RM3-TH/06-6}\\
{\tt TUM-HEP-629/06}\\
\end{tabular}
\end{flushright}

\vspace{0.5cm}
\begin{center}

\LARGE{\bf Exploring twisted mass Lattice QCD
\\ \vspace{0.2cm} with the Clover term}

\vspace{0.5cm}
{\sc \large{D.~Be\'cirevi\'c$^{\,a}$, Ph.~Boucaud$^{\,a}$,
V.~Lubicz$^{\,b,c}$, G.~Martinelli$^{\,d}$, F.~Mescia$^{\,e}$,
S.~Simula$^{\,c}$, C.~Tarantino$^{\,f}$}}

\vspace{0.5cm}
\normalsize{\sl 
$^a$Laboratoire de Physique Th\'eorique (B\^at.~210), Universit\'e
    de Paris XI, \\ Centre d'Orsay, 91405 Orsay-Cedex, France\\
$^b$Dipartimento di Fisica, Universit\`a di Roma Tre, Via della Vasca Navale
84, I-00146 Rome, Italy \\
$^c$INFN, Sezione di Roma III, Via della Vasca Navale 84, I-00146 Roma,
Italy\\
$^d$Dipartimento di Fisica, Universit\`a di Roma ``La Sapienza'', 
    and INFN, Sezione di Roma, P.le A.~Moro 2, I-00185 Rome, Italy\\
$^e$INFN, Laboratori Nazionali di Frascati, Via E. Fermi 40, I-00044
Frascati, Italy\\
$^f$Physik Depart., Technische Universit\"at M\"unchen, D-85748
Garching, Germany
}
\end{center}

\vspace{0.2cm}
\begin{abstract}
It has been shown that in the twisted mass formulation of Lattice QCD at
maximal twist large cutoff effects are generated when the quark mass becomes
of ${\cal O}(a \Lambda_{QCD}^2)$. In general, these effects can be
suppressed in two ways: either by choosing the critical quark mass in an
``optimal way", or by adding the Clover term to the twisted action. We
investigate the second option by performing a quenched lattice QCD
simulation with twisted Clover fermions and pion masses as low as 280 MeV.
We show that the Clover term is indeed efficient in reducing the large
cutoff effects. In particular, the so-called bending phenomenon observed in
the determination of the pion decay constant is cured in this way. In
addition, by using maximally twisted Clover fermions, we provide a
non-perturbative determination of the vector current renormalization
constant $Z_V$ as well as of the non-perturbatively renormalized light quark
masses. Finally, we calculate the connected contribution to the
charged-neutral pseudoscalar meson mass splitting, finding that the
introduction of the Clover term in the twisted action is also beneficial, in
the quenched approximation, in reducing cutoff effects related to the
isospin symmetry breaking at finite lattice spacing.
\end{abstract}

\newpage
\section{Introduction}
The twisted mass formulation of Lattice QCD (tmLQCD)~\cite{tmlqcd,shindler}
offers a number of interesting advantages. The absence of real unphysical
zero modes, which affect the standard Wilson discretization of the fermionic
action, prevents the appearance in this framework of the exceptional
configurations in quenched and partially-quenched numerical simulations.
This allows to investigate QCD dynamics with much lighter valence quarks. In
addition, twisting the Wilson term in the action (in the physical basis)
considerably simplifies the renormalization pattern of several local
operators. Notable examples in this respect are the determination of the
pion decay constant, which can be performed without introducing any
renormalization factor~\cite{tmlqcd}, and of the kaon $B_K$ parameter, which
does not require subtraction of wrong chirality
operators~\cite{tmlqcd,vlad}.\footnote{For an alternative approach which
avoids this subtraction even when working with ordinary Wilson fermions see
Refs.~\cite{Becirevic:2000cy,Becirevic:2004aj}.} Furthermore, tmLQCD with a
maximal twist angle ($\alpha=\pi/2$), denoted as maximally twisted mass LQCD
(Mtm-LQCD), ensures automatic (or almost automatic) ${\cal
O}(a)$-improvement for the physically interesting quantities computed on the
lattice~\cite{mtm}.

The drawback, however, is that with tmLQCD both parity and flavor symmetries
are explicitly broken at finite lattice spacing. The breaking of parity
generates a mixing with wrong parity states which may affect, in some
relevant cases, the large time behavior of lattice correlation functions.
The breaking of flavor symmetry is responsible for generating a mass
splitting between otherwise degenerate isospin hadronic partners. Although
the breaking effects are parametrically of ${\cal O}(a^2)$ at maximal twist,
this splitting has been found to be numerically significant in the case of
the neutral and charged pseudoscalar mesons~\cite{pisplit}.

Another important issue of practical interest for Mtm-LQCD is the presence
of large cutoff effects which are generated when the quark mass $m_q$
becomes of ${\cal O}(a \Lambda_{QCD}^2)$ or smaller. This phenomenon has
been the subject of recent numerical and theoretical investigations. From
the point of view of the Symanzik expansion, these large cutoff effects have
been studied in Ref.~\cite{ird}. In correlation functions they are related
to multiple insertions of the dimension-five operator $\mc{L}_5$ entering
the Symanzik effective Lagrangian for Mtm-LQCD close to the continuum limit.
Since, at maximal twist, $\mc{L}_5$ has the quantum numbers of the neutral
pion, its insertions can create from the vacuum Goldstone bosons with zero
momentum, thus generating discretization effects which are proportional, at
the leading order, to $(a / M_\pi^2)^2 \sim (a/m_q)^2$. In this situation,
the value of the quark mass is limited to the region $m_q\simge a
\Lambda_{QCD}^2$. Being proportional to $(a/m_q)^2$, such lattice artifacts
have been also called ``infra-red divergent" (IRD) discretization effects.
Note, however, that they actually do not represent any infra-red divergence,
since the chiral limit can be taken only after the continuum extrapolation
has been performed.

A further step forward in understanding the IRD cutoff effects has been made
in Ref.~\cite{ird-chpt}, where it is shown that their contribution to the
Symanzik expansion can be explicitly resummed using twisted-mass chiral
perturbation theory (tm$\chi$PT)~\cite{tmchpt}. For a simple quantity like
the pion decay constant $f_\pi$, the leading order result of tm$\chi$PT
reads
\be
f_\pi = f ~ \frac{\mu}{\sqrt{\mu^2 + \delta m^2}} \simeq f ~ \left[ 1 -
\frac{1}{2} \left(\frac{\delta m}{\mu}\right)^2 + ~ ... \right]~,
\label{fpi-ird}
\ee
where $f$ is the pion decay constant in the chiral limit and $\mu$ and
$\delta m$ represent the (renormalized) twisted and non-twisted quark masses
respectively. Since at maximal twist $\delta m$ is a quantity of $\mc{O}(a)$
introduced by a discretization error in the determination of the
critical mass, Eq.~(\ref{fpi-ird}) shows that the deviation of $f_\pi$ from
its value in the chiral limit has the form of an IRD cutoff effect. The
leading correction is proportional to $(\delta m/\mu)^2 \sim (a/M_\pi^2)^2$
and its contribution becomes sizable when the twisted mass $\mu$ is small
and comparable to $\delta m$, i.e., $\mu \sim \mc{O}(a \Lambda_{QCD}^2)$.

The requirement $\mu \simge a \Lambda_{QCD}^2$ is in practice rather
restrictive. For a typical inverse lattice spacing $a^{-1} \simeq 2$ GeV, it
implies pion masses $M_\pi \simge 500$ MeV. In order to allow for
simulations with smaller values of the quark masses (down to $\mu \simge a^2
\Lambda_{QCD}^3$), and keep discretization effects under control, two
strategies have been proposed~\cite{ird}. Either one uses an ``optimal"
determination of the critical mass $m_{cr}$ or the (twisted) Clover term is
introduced in the quark action. In the latter case, the coefficient $c_{SW}$
has to be fixed to the same value that guarantees on-shell
$\mc{O}(a)$-improvement with ordinary Wilson fermions.

The idea behind the first strategy, namely the optimal determination of the
critical mass is to tune the $\mc{O}(a)$ contribution to $m_{cr}$ in such a
way that the dimension-five operator in the Symanzik effective Lagrangian
has vanishing matrix element between the vacuum and the single pion state,
i.e. $\langle 0\vert \mc{L}_5\vert \pi^0\rangle =0$~\cite{ird}. Since the
IRD cutoff effects have been shown to be proportional at the leading order
to the previous matrix element, this procedure guarantees their suppression
and allows much smaller values of the quark masses to be safely reached in
the simulation. In practice, this procedure corresponds to fix the value of
the critical mass by requiring the vanishing of the PCAC quark mass in the
twisted basis, as it was anticipated in Refs.~\cite{sharpe,aoki}. In the
following, we will denote as $m_{cr}^{\rm opt}$ this ``optimal"
determination of the critical mass.

The numerical studies reported in Refs.~\cite{XLF,Abdel-Rehim} show that
such a strategy is indeed efficient. In particular, in agreement with
theoretical expectations, it helps in suppressing the pronounced bending
behaviour of the pion decay constant, i.e., the appearance of large lattice
artifacts in this quantity when the quark mass becomes of
$\mc{O}(a\Lambda_{QCD}^2)$ or smaller. Note, however, that one has to make a
preliminary numerical study, with twisted fermions, to determine the optimal
value of the critical mass. Such a determination may be computationally
expensive, particularly in the case of unquenched simulations.

In this paper we explore the second strategy of Ref.~\cite{ird}, namely the
simulation of Mtm-LQCD with the inclusion of the Clover term in the action.
In this case, in the Symanzik expansion of the lattice effective Lagrangian
at $\mc{O}(a)$ only a term proportional to the square of the twisted quark
mass survives~\cite{Sharpe:2004ps}, that is $\mc{L}_5=\mc{O}(a\mu^2)=
\mc{O}(aM_\pi^4)$. Thus, the introduction of the Clover term is expected to
be efficient in cancelling the leading order IRD cutoff effects as much as
the tuning of the critical mass in the twisted Wilson case.

From the previous discussion it should be clear that the Mtm-Clover approach
does not require any specific prescription for the determination of the
critical mass. We show here that this is indeed the case. We have performed
two numerical simulations with Mtm-Clover fermions and with a critical quark
mass determined in the non-twisted theory either by requiring the vanishing
of the pion mass in the chiral limit (denoted in the following as
$m_{cr}^{\rm pion}$) or by imposing the vanishing of the PCAC mass
($m_{cr}^{\rm PCAC}$). In both cases, we do not observe any bending
phenomenon in the pion decay constant, even at values of quark masses
smaller than $\mc{O}(a\Lambda_{QCD}^2)$. We also compare our results for the
pion mass and decay constant with those obtained by the $\XLF$
Collaboration~\cite{XLF} using Mtm-Wilson fermions with the optimal choice
of the critical mass, finding a very good agreement. The appealing feature
of the Mtm-Clover approach is that the determination of the critical mass
does not require any specific tuning and can be obtained from a single
simulation performed with non-twisted Clover fermions.

Besides investigating the effectiveness of Mtm-Clover fermions in
suppressing the IRD cutoff effects, we also present a non-perturbative
determination of the local vector current renormalization constant $Z_V$
which renormalizes, in Mtm-LQCD, the local axial-vector current (in the
physical basis). This constant is expected to be identical, up to
$\mc{O}(a^2)$ effects, to the value determined by using non-twisted Clover
fermions. Indeed, we find a very good agreement between our result and the
non-twisted determination of $Z_V$ performed in Ref.~\cite{ri-mom} with the
non-perturbative RI-MOM method.

In addition, we present a calculation of the strange and of the average
up/down quark masses using the value of the quark mass renormalization
constant determined non-perturbatively in Ref.~\cite{ri-mom}. With respect
to previous quenched determinations of light quark masses, here we benefit
from the use of Mtm-LQCD with the Clover term to probe ever lower values of
the quark masses, and thereby reduce the errors due to the chiral
extrapolation when determining the light up/down quark masses on the
lattice.

Finally, we investigate the isospin symmetry breaking effects, induced at
finite lattice spacing by the tmLQCD action, by studying the contribution of
connected diagrams to the charged-neutral pion mass splitting. This
connected contribution can be interpreted as the charged-neutral mass
splitting between pseudoscalar mesons composed by quarks and antiquarks
belonging to different, though degenerate, isospin doublets. By comparing
the results obtained with Mtm-Wilson and Mtm-Clover fermions, we find that
the introduction of the Clover term is beneficial also in reducing these
cutoff effects, at least for the pure connected contribution considered in
this study and in the quenched approximation.

It is also worth noting that the same ${\cal O}(a^2)$ operator that provides
the leading contribution to the pion mass splitting in Mtm-LQCD also affects
the determination of both the charged and the neutral pion masses when
working with ordinary Wilson/Clover fermions. In this respect, therefore,
the use of Mtm-LQCD does not introduce any additional uncertainty.

The plan of the paper is as follows. In Section~\ref{sec:fpi} we give the
details of the lattice simulation and present our results for the
pseudoscalar meson mass and decay constant; we show the effectiveness of the
introduction of the Clover term in curing the bending phenomenon. We present
the determination of the renormalization constant $Z_V$ in
Section~\ref{sec:zv} and the calculation of the light quark masses in
Section ~\ref{sec:qm}. In Section~\ref{sec:neutral} we evaluate the
contribution of connected diagrams to the charged-neutral pion mass
splitting and compare the results with those obtained by using Mtm-Wilson
fermions with the optimal choice of the critical mass. Our conclusions are
summarized in Section~\ref{sec:conclusions}.

\section{Pseudoscalar meson masses and decay constants with the Mtm-Clover
action\label{sec:fpi}}

The numerical results presented in this study have been obtained from a set
of 300 gauge configurations generated, in the quenched approximation, with
the standard Wilson gauge action at $\beta = 6.0$, which corresponds to an
inverse lattice spacing $a^{-1} \simeq 2.0 ~ {\rm GeV}$. The lattice volume
is $16^3 \times 32$ so that the physical length of the lattice is about 1.6
fm in the spatial directions. Using this set of gauge configurations, quark
propagators have been computed by implementing three different choices for
the quark action:
\begin{itemize}
\item[a)] A simulation with Mtm-Clover fermions and the critical mass
determined from the PCAC relation in the non-twisted theory, $m_{cr}=m_{cr}
^{\rm PCAC}$. The value of the critical hopping parameter, $\kappa_{cr}^{\rm
PCAC} =0.135217$, corresponding to $am_{cr}^{\rm PCAC} \simeq -0.3022$, has
been obtained by using the results of a previous simulation done by our
collaboration with standard Clover fermions at the same value of the gauge
coupling on the volume $24^3\times 64$~\cite{ri-mom}.

\item[b)] A simulation with Mtm-Clover fermions and the critical mass
determined from the vanishing of the pion mass in the non-twisted theory,
$m_{cr}=m_{cr}^{\rm pion}$. The value of the critical hopping parameter,
$\kappa_{cr}^{\rm pion} =0.135293$, corresponds to $a m_{cr}^{\rm pion}
\simeq -0.3043$ and has been determined from the same set of non-twisted
Clover data used to determine the value of $m_{cr}^{\rm PCAC}$ in simulation
a).

\item[c)] A simulation with Mtm-Wilson fermions with the critical mass
determined in the optimal way, that is $m_{cr}=m_{cr}^{\rm opt}$. We used
the value $\kappa_{cr}^{\rm opt} = 0.157409$ obtained by the $\XLF$
Collaboration from a simulation at $\beta=6.0$ on the volume $16^3\times
32$~\cite{XLF}.
\end{itemize}

In simulations a) and b) with Mtm-Clover fermions the coefficient of the
Clover term has been fixed to the value determined non-perturbatively in
Ref.~\cite{luscher-csw}, namely $c_{SW}=1.769$.

In all three sets of simulations we have computed quark propagators with the
same 9 values of the bare twisted quark mass $a\mu$ (cf.
Table~\ref{tab:masses}), which are equal to those used in Ref.~\cite{XLF}.
The range of chosen $a\mu$ values covers pion masses between approximately
280 MeV and 1.1 GeV. In order to invert the Dirac operator for such a broad
set of quark masses we employed the multiple mass solver
algorithm~\cite{MMS}.

To examine the impact of cutoff effects when working with maximally twisted
fermions at small values of the quark mass, we investigate whether or not
the inclusion of the Clover term is efficient in curing the bending
phenomenon observed in the determination of the pion decay constant with
Mtm-Wilson fermions, when the critical mass is determined in a non-optimal
way.

Pseudoscalar meson masses and decay constants have been evaluated by
studying the large time behavior of the following two-point correlation
functions:
 \be
 C_{PP}(t) \equiv \sum_{\vec{x}} \langle P_5(t, \vec{x}) P_5^\dagger(0) 
    \rangle \quad , \quad
 C_{AP}(t) \equiv \sum_{\vec{x}} \langle A_0(t, \vec{x}) P_5^\dagger(0) 
    \rangle ~ ,
    \label{eq:correlators}
 \ee
where $P_5 = \bar{u} \gamma_5 d$ and $A_\mu = \bar{u} \gamma_\mu\gamma_5 d$.
Throughout this paper we choose to  work in the physical basis and use
mass-degenerate $u$ and $d$ quarks.

Using the completeness relation and considering large enough time
separations, one gets
\beq
C_{PP}(t) ~ & _{\overrightarrow{t\gg 1}} & ~ \frac{Z_{PS}}{M_P} ~
e^{-M_P T/2} \, \cosh\left(M_P(T/2-t)\right)  ~ , \nn \\
C_{AP}(t) ~ & _{\overrightarrow{t \gg 1}} & ~ \frac{f_P M_P}{Z_V} ~
\frac{\sqrt{Z_{PS}}}{M_P} ~ e^{-M_P T/2} \,
\sinh\left(M_P(T/2-t)\right)~ , 
\label{eq:CApi}
\eeq
where $M_P$ is the mass of the lightest pseudoscalar meson state,
$\sqrt{Z_{PS}} = \vert \langle 0 | P_5(0) | \pi \rangle\vert$ and $f_P M_P =
Z_V \langle 0 | A_0(0) | \pi \rangle$. $Z_V$ is the renormalization constant
of the local axial-vector current with Mtm-LQCD, which corresponds to the
renormalization constant of the local vector current with the non-twisted
quark action~\cite{mtm}. The pseudoscalar decay constant can therefore be
extracted by studying the ratio $C_{AP}(t)/C_{PP}(t)$ at large time
distances,
\be
f_P = Z_V \frac{\sqrt{Z_{PS}}}{M_P} ~ \left[ \frac{C_{AP}(t)}{C_{PP}(t)}\,
\coth\left(M_P(T/2-t)\right)\right]_{t \gg 1} ~ .
\label{eq:fpi_direct}
\ee

An alternative determination of the decay constant~\cite{tmlqcd}, which does
not require the introduction of any renormalization constant, can be
obtained by exploiting the consequences of the axial-vector Ward identity
(AWI) for Mtm-LQCD. Up to $\mc{O}(a^2)$ correction, the relevant identity
reads
 \be
Z_V\, \sum_{\vec{x}} \langle \partial_0 A_0(t,\vec{x}) P_5^\dagger(0)\rangle
= 2 \mu\,\sum_{\vec{x}} \langle P_5(t,\vec{x}) P_5^\dagger(0)\rangle  ~,
 \label{eq:awi}
 \ee
where $\mu$ is the bare twisted
mass. Combining the above relation with Eqs.~(\ref{eq:CApi}) one gets
 \be
    f_P = 2 \mu \frac{\sqrt{Z_{PS}}}{M_P^2} ~ ,
    \label{eq:fpi_indirect}
 \ee
where both $Z_{PS}$ and $M_P$ can be determined from the correlation
function $C_{PP}(t)$ of Eq.~(\ref{eq:CApi}). We have evaluated the
pseudoscalar meson decay constant $f_P$ using both the ``direct" method
(\ref{eq:fpi_direct}) and the ``indirect" one (\ref{eq:fpi_indirect}),
obtaining always a very good agreement within the statistical
errors.\footnote{This agreement is further improved if the factor $M_P^2$ in
the denominator of Eq.~(\ref{eq:fpi_indirect}) is replaced with $M_P\,
\sinh(a M_P)/a$, where the sinh is originated from the discretized symmetric
lattice version of the derivative in Eq.~(\ref{eq:awi}). Our results from
Eq.~(\ref{eq:fpi_indirect}) have been obtained with this replacement always
implemented.} For the direct method, we used the non-perturbative RI-MOM
determination $Z_V^{(Clover)}=0.772$~\cite{ri-mom} and the one-loop boosted
perturbative estimate $Z_V^{(Wilson)}=0.63$~\cite{Capitani,lm} in the
Mtm-Clover and Wilson case respectively. In what follows, we only quote the
results for $f_P$ obtained from the indirect method~(\ref{eq:fpi_indirect}).

\begin{figure}[p]
\vspace{-1.35cm}
\begin{center}
\includegraphics[bb=0cm 17cm 30cm 30cm, scale=0.65]{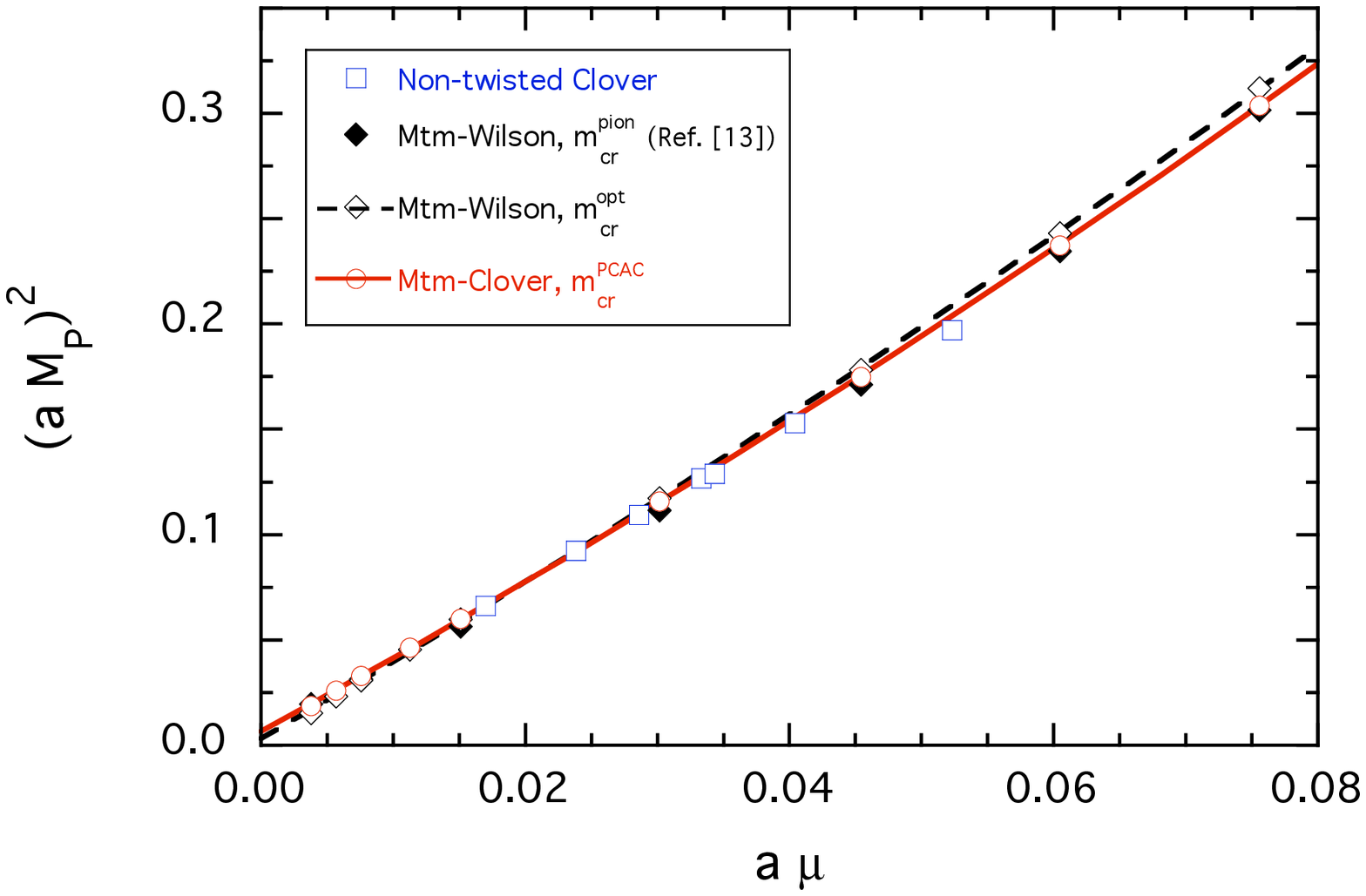}
\end{center}
\caption{\small\sl Values of the pion mass squared obtained with different
${\cal O}(a)$-improved actions, as specified in the legend and in the text.
The solid and dashed lines represent quadratic fits in $a \mu$.}
\label{fig:pimass}
\end{figure}

\begin{figure}[p]
\vspace{-1.35cm}
\begin{center}
\includegraphics[bb=0cm 17cm 30cm 30cm, scale=0.65]{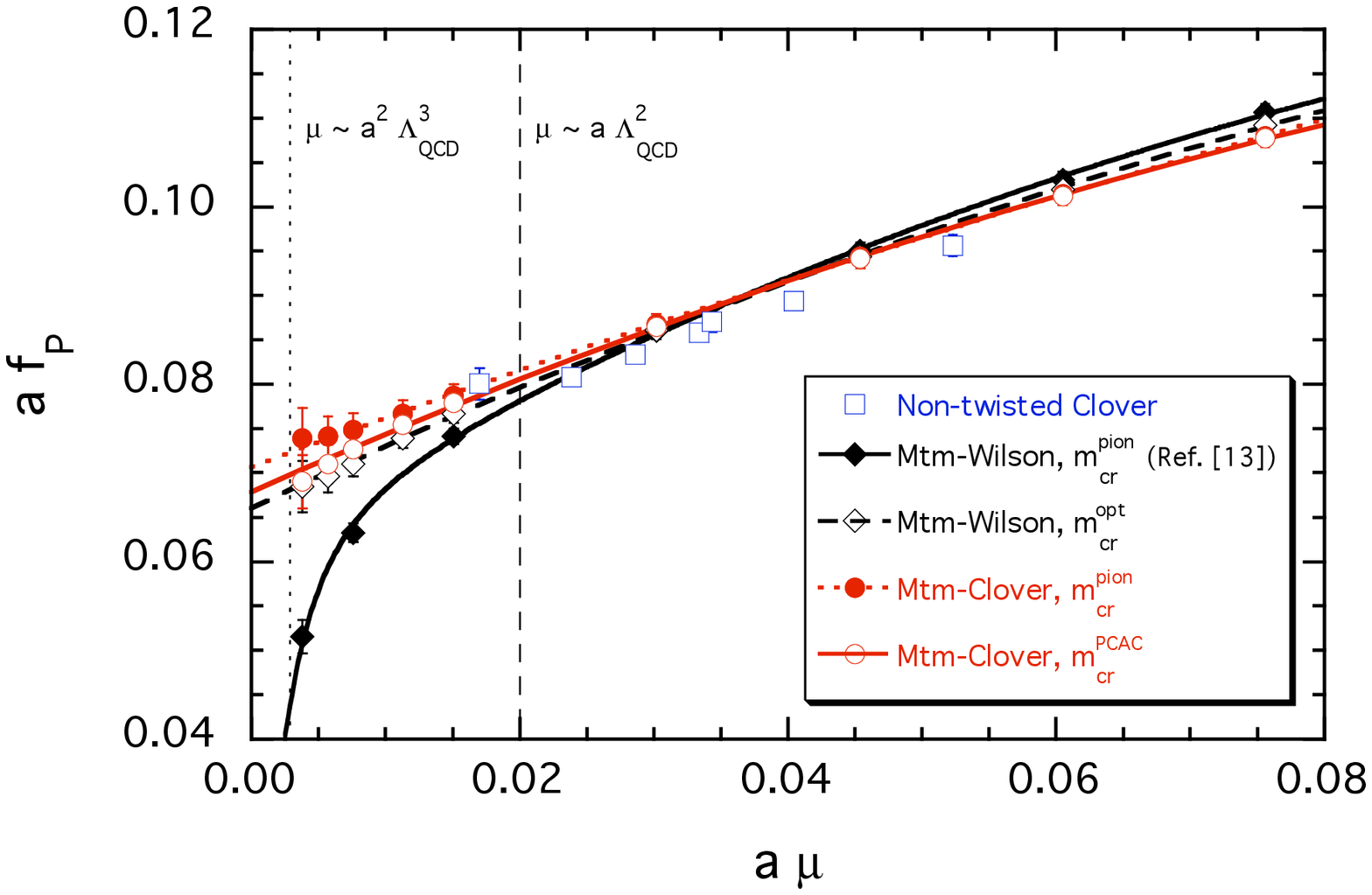}
\end{center}
\caption{\small\sl Same as in Fig.~\ref{fig:pimass}, but for the
pseudoscalar decay constant. The lines represent quadratic fit in $a\mu$
except for the results obtained with Mtm-Wilson fermions and the
$m_{cr}^{\rm pion}$ determination of the critical mass. In this case, the
fit also takes into account the tm$\chi$PT prediction expressed by
Eq.~(\ref{fpi-ird}).}
\label{fig:fpi}
\end{figure}

Our result for the pseudoscalar meson masses and decay constants obtained
from simulations a), b) and c) are collected in Table~\ref{tab:masses}.
\begin{table}[t]
\vspace{0.5cm}
\begin{center}
\small{
\begin{tabular}{||c||c|c||c|c||c|c||}
 \cline{2-7} \multicolumn{1}{c||}{{\phantom{\huge{l}}}\raisebox{-.2cm}{
\phantom{\Huge{j}}}} & 
\multicolumn{2}{c||}{a) Mtm-Clover, $m_{cr}^{\rm PCAC}$} &
\multicolumn{2}{c||}{b) Mtm-Clover, $m_{cr}^{\rm pion}$} &
\multicolumn{2}{c||}{c) Mtm-Wilson, $m_{cr}^{\rm opt}$}
\\  \hline
 $a \mu$   & $a M_P$  & $a f_P$  & $ a M_P$ & $a f_P$  & $ a M_P$ & $a f_P$
\\ \hline \hline
 $0.0038$  & $0.1359~(66)$ & $0.0690~(30)$ & $0.1308~(70)$ & $0.0739~(34)$ 
 & $0.1233~(68)$ & $0.0684~(29)$
\\ \hline
 $0.0057$  & $0.1607~(52)$ & $0.0710~(22)$ & $0.1571~(51)$ & $0.0741~(23)$ 
 & $0.1515~(52)$ & $0.0696~(18)$
\\ \hline
 $0.0076$  & $0.1811~(46)$ & $0.0727~(18)$ & $0.1787~(47)$ & $0.0748~(19)$ 
 & $0.1751~(44)$ & $0.0710~(14)$
\\ \hline
 $0.0113$  & $0.2150~(39)$ & $0.0754~(15)$ & $0.2137~(39)$ & $0.0767~(15)$ 
 & $0.2127~(36)$ & $0.0739~(11)$
\\ \hline
 $0.0151$  & $0.2449~(35)$ & $0.0779~(13)$ & $0.2444~(35)$ & $0.0787~(13)$ 
 & $0.2445~(31)$ & $0.0767~(10)$
\\ \hline
 $0.0302$  & $0.3401~(26)$ & $0.0865~(11)$ & $0.3404~(26)$ & $0.0868~(11)$ 
 & $0.3425~(22)$ & $0.0861~(09)$
\\ \hline
 $0.0454$  & $0.4182~(22)$ & $0.0942~(11)$ & $0.4187~(22)$ & $0.0944~(11)$ 
 & $0.4223~(18)$ & $0.0944~(08)$
\\ \hline
 $0.0605$  & $0.4873~(20)$ & $0.1012~(10)$ & $0.4879~(20)$ & $0.1014~(10)$ 
 & $0.4930~(16)$ & $0.1019~(08)$
\\ \hline
 $0.0756$  & $0.5512~(17)$ & $0.1077~(10)$ & $0.5519~(18)$ & $0.1079~(10)$ 
 & $0.5585~(14)$ & $0.1092~(08)$
\\ \hline
 \hline
\end{tabular}}
\end{center}
\vspace{-0.3cm}
\caption{\small\sl Values of the bare twisted quark mass $a\mu$, of the
pseudoscalar meson mass $aM_P$ and of the pseudoscalar meson decay constant
$af_P$ from the three sets of simulations performed in this study. The
quoted errors are purely statistical.}
\label{tab:masses}
\end{table}
These results are also shown in Figs.~\ref{fig:pimass} and \ref{fig:fpi}, as
a function of the bare quark mass $a\mu$.
The Mtm-Wilson results from our simulation c) agree well with those
presented in Ref.~\cite{XLF}. For comparison, we also show in these plots
the results obtained in~\cite{XLF} by employing Mtm-Wilson fermions with the
non-optimal choice of the critical mass, as well as the results obtained
with non-twisted Clover fermions (at $\beta=6.0$) using data produced by our
collaboration for the studies described in Refs.~\cite{ri-mom,Clover}. In
the non-twisted case, the quark mass $a \mu$ in Figs.~\ref{fig:pimass} and
\ref{fig:fpi} represents the bare AWI quark mass $a m^{\rm
AWI}$~\cite{Clover} multiplied by the axial-vector renormalization constant
$Z_A$. Indeed, this is the quantity that renormalizes with the same
renormalization constant $Z_\mu=1/Z_P$ of the bare twisted mass $a \mu$ in
the twisted formulation of LQCD.\footnote{It should be also mentioned that
the bare twisted quark mass $a \mu$ with Mtm-Clover fermions renormalizes
differently from its counterpart with Mtm-Wilson fermions. However, the
ratio of the corresponding renormalization constants, $Z_P^{(Clover)} /
Z_P^{(Wilson)}$, is very close to unity (it is $0.97$ when using the
one-loop boosted perturbative expressions).}

From the plots in Figs.~\ref{fig:pimass} and \ref{fig:fpi} we observe that:
\begin{itemize}
\item There is a very good agreement among the results for the pseudoscalar
meson masses calculated using various quark actions, all of which are ${\cal
O}(a)$-improved. This may signal the smallness of discretization effects on
this quantity at the value of the lattice spacing considered in the present
study. Further investigations at different lattice spacings could better
clarify this point.

\item IRD cutoff effects are clearly visible in the pseudoscalar meson decay
constant at low quark masses ($a\mu \simle 0.02 \sim a^2 \Lambda_{QCD}^2$)
when computed by using Mtm-Wilson fermions with the $m_{cr}^{\rm pion}$
determination of the critical mass. This is the bending phenomenon observed
in Ref.~\cite{XLF}. As shown in Fig.~\ref{fig:fpi}, these effects are
strongly reduced either by choosing the optimal value of $m_{cr}$, or by
introducing the Clover term in the quark action. The reduction of IRD
effects is almost the same within the two approaches. This is the main
result of the present study.

\item As shown in Fig.~\ref{fig:pimass}, IRD cutoff effects are negligible
in the case of the pseudoscalar meson mass. This is consistent with the
observation that IRD cutoff effects in the determination of this mass are
softened by an additional factor of $M_P^2$~\cite{ird}.

\item The results for the pseudoscalar meson masses obtained with the
Mtm-Clover action and the $m_{cr}^{\rm pion}$ determination of the critical
mass are not shown in Fig.~\ref{fig:pimass}, because they are practically
indistinguishable from those obtained with $m_{cr}^{\rm PCAC}$ (cf.
Table~\ref{tab:masses}). In the case of the decay constant, instead, the two
Mtm-Clover determinations with different choices of $m_{cr}$ show a small
spread at very low quark masses. This may signal the contribution of
higher-order IRD cutoff effects that we cannot cure.
\end{itemize}

For illustrative purposes we quote the results for the pion decay constants
obtained by quadratically extrapolating the lattice data to the physical
pion mass. We obtain $f_\pi = \{138\,(8),~ 144\,(9),~138\,(8)\}$ MeV in
simulations a), b) and c) respectively, to be compared with the experimental
value $f_\pi=132$ MeV.

\section{More numerical results\label{sec:other}}

\subsection{The renormalization constant $Z_V$\label{sec:zv}}
A non-perturbative determination of the renormalization constant
$Z_V$~\footnote{In this paper we follow the standard practice to denote with
$Z_V$ the renormalization constant of the axial-vector current in Mtm-LQCD.
The reason is that the axial-vector current corresponds to the vector
current in the twisted basis, which also implies that $Z_V$ coincides with
the vector current renormalization constant of the standard non-twisted
theory.} can be obtained using the two-point AWI of Eq.~(\ref{eq:awi}) as
\be
\label{eq:ZV}
Z_V^{-1} =
\frac{\left(C_{AP}(t+1)-C_{AP}(t-1)\right)/2}{2\,\mu\,C_{PP}(t)}~,
\ee
where in the numerator we used the symmetric ($\mc{O}(a)$-improved) version
of the lattice time derivative.

\begin{figure}[t]
\begin{center}
\includegraphics[bb=0cm 20.5cm 30cm 30cm, scale=0.75]{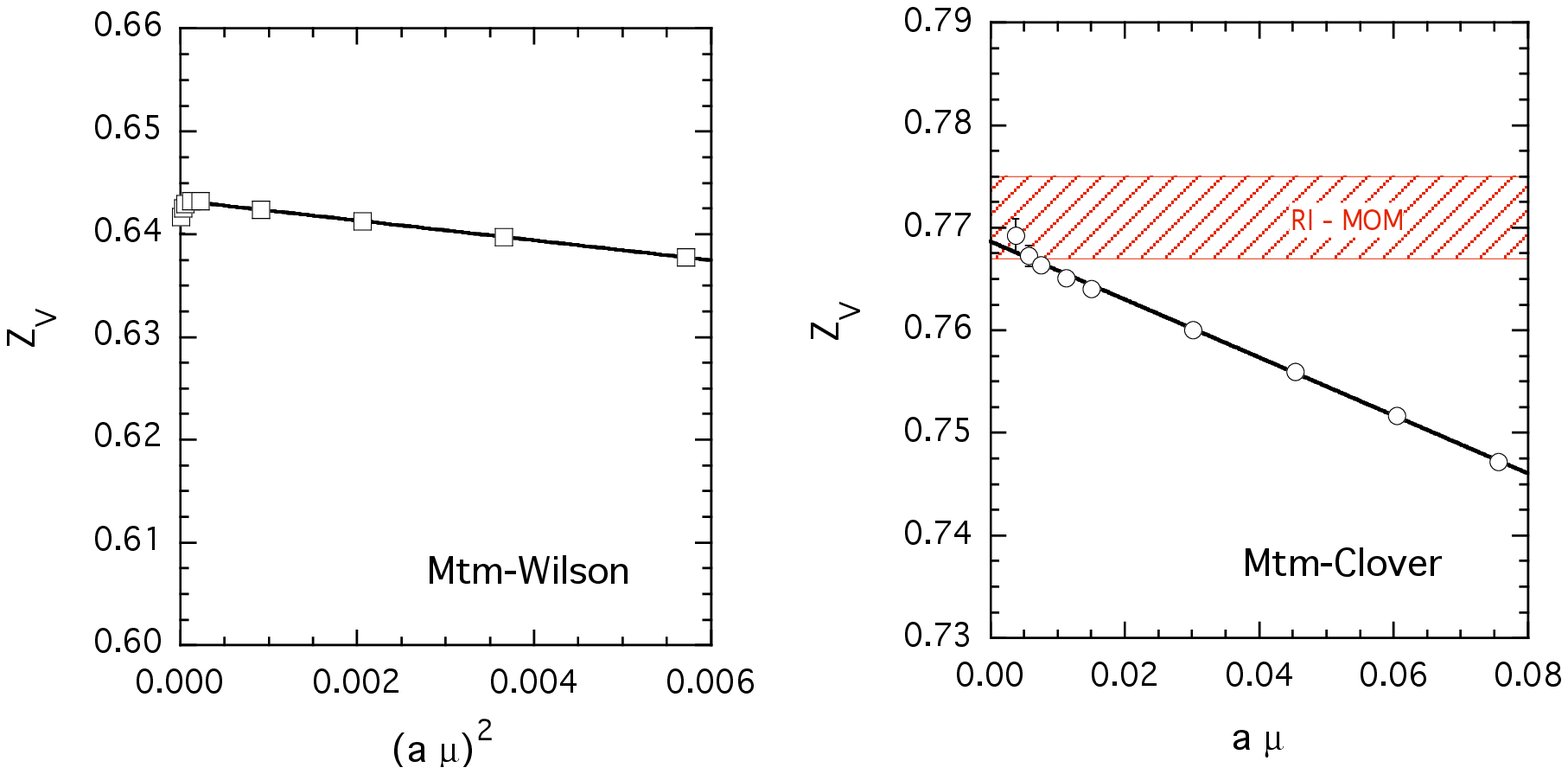}
\end{center}
\caption{\small\sl Values of the renormalization constant $Z_V$ as a
function of the bare twisted quark mass as obtained by using the Mtm-Wilson
action (left) and the Mtm-Clover action (right). Note that the square mass
$(a\mu)^2$ and the mass $(a\mu)$ are plotted respectively on the x-axis in
the two cases. Solid lines are fits to the lattice data. The hatched area in
the right plot shows the non-perturbative RI-MOM estimate of $Z_V$ obtained
from non-twisted Clover fermions in Ref.~\cite{ri-mom}.}
\label{fig:ZV}
\end{figure}

We have computed $Z_V$ from Eq.~(\ref{eq:ZV}) using both the Wilson and the
Clover action at maximal twist. In the Clover case, the results obtained
from simulations a) and b), which only differ by the choice of the critical
mass, are in good agreement within the statistical errors. For this reason,
only data from simulation a) and c) will be discussed below. Our results for
$Z_V$ are shown in Fig.~\ref{fig:ZV} as a function of the bare twisted quark
mass.

The dependence of $Z_V$ on the quark mass is induced by discretization
effects, which are predicted to be of $\mc{O}(a^2)$ at maximal twist. Thus
one can expect both a quadratic dependence on the quark mass, due to
contributions of $\mc{O}(a^2 \mu^2)$, and a linear dependence induced by
terms of $\mc{O}(a^2 \Lambda_{QCD}\,\mu)$.\footnote{We thank R. Frezzotti
and G. Rossi for having drawn our attention to this point.} From the results
shown in Fig.~\ref{fig:ZV}, it can be clearly seen that terms of $\mc{O}(a^2
\Lambda_{QCD}\,\mu)$ are dominant in the Mtm-Clover case, while the
quadratic contributions of $\mc{O}(a^2 \mu^2)$ are larger in the Wilson
case, except that at very small values of the quark mass.

In both the Wilson and Clover cases, we can correct for mass-dependent
discretization effects by extrapolating the results for $Z_V$ to the chiral
limit. In this way, we obtained the estimates
\be
Z_V^{(Wilson)} = 0.643(1) \qquad, \qquad Z_V^{(Clover)} = 0.768(1)~.
\label{eq:zvres}
\ee
Our result for $Z_V^{(Wilson)}$ agrees with the one reported in
Ref.~\cite{XLFscaling}, namely $Z_V^{(Wilson)}=0.6424(4)$.

The determinations in Eq.~(\ref{eq:zvres}) can be also compared to the
estimates of $Z_V$ obtained in the corresponding non-twisted theories (for
which $\mu=0$), since renormalization constants are independent of the quark
mass. For the standard Clover action, non-perturbative determinations of
$Z_V$ have been obtained using several approaches. In Ref.~\cite{ri-mom}, by
using the RI-MOM method the estimate $Z_V^{(Clover)}=0.772(2)(2)$ has been
obtained, in good agreement with our result in Eq.~(\ref{eq:zvres}) (this
comparison is also illustrated in Fig.~\ref{fig:ZV}). Similar estimates have
been obtained with the other approaches~\cite{zvalpha,zvlanl,zvqcdsf}. The
determinations of $Z_V$ with standard Wilson fermions are affected, instead,
by much larger uncertainties, since the action is not improved at
$\mc{O}(a)$. Typically, the non-perturbative estimates lie in the range
$Z_V^{(Wilson)} \simeq 0.6 \div 0.8$~\cite{zvqcdsf,zvrap,zvcppacs}. At
variance with these estimates, however, we remind that our determination in
Eq.~(\ref{eq:zvres}) is improved at $\mc{O}(a)$. Another interesting
comparison is provided by the estimates $Z_V^{(Wilson)} \simeq 0.63$ and
$Z_V^{(Clover)} \simeq 0.81$ obtained using one-loop tadpole-improved
boosted perturbation theory~\cite{Capitani,lm}.\footnote{We choose
the boosted coupling obtained by inverting the perturbative series of the
logarithm of the plaquette,
\be
\ln \langle P \rangle = -\frac{1}{3} \tilde g^2(3.40/a)\left[1-1.1905\,
\frac{\tilde g^2(3.40/a)}{4\pi}\right]
\ee}

\subsection{Light quark masses\label{sec:qm}}
In the quenched approximation, one of the main advantages of tmLQCD is the
possibility to perform simulations at small values of the quark masses,
without encountering the problem of exceptional configurations. This can be
very beneficiary for the extrapolations towards the chiral limit. A notable
example, in this respect, is the calculation of the light quark masses and,
in particular, of the average up/down quark mass.

Moreover, the determination of quark masses with Mtm-LQCD is simpler than
with the standard Wilson and/or Clover actions. To emphasize this point we
remind the reader that the renormalized quark mass in Mtm-LQCD is obtained
from the twisted mass parameter $\mu$ as $\hat m_q = Z_\mu \mu$, with $Z_\mu
= 1/Z_P$. Thus $\mu$ plays a role similar to the bare $m^{\rm AWI}$ in the
standard, non-twisted, case. But contrary to $m^{\rm AWI}$, which is
obtained from the ratio of the correlation functions and thus suffers from 
both statistical and systematic errors, the bare twisted mass $\mu$ is known
without uncertainties. In addition to the automatic ${\cal
O}(a)$-improvement of the results obtained by using either the Mtm-Wilson or
the Mtm-Clover action, this makes the calculation of the light quark mass
particularly attractive.

To exemplify the benefits of using the Mtm-LQCD we now apply the same
strategy explained in detail in Ref.~\cite{Clover}, and determine the values
of the strange and the average up/down quark masses. Concerning the
renormalization constant, in the Mtm-Clover case we use the non-perturbative
estimate $Z_P^{\rm RI/MOM}(2\gev)=0.525$~\cite{ri-mom}, which corresponds
to $Z_P^{\overline{MS}}(2\gev)=0.624$. In the Mtm-Wilson case, instead, we
will use the one-loop boosted perturbative result, namely
$Z_P^{\overline{MS}}(2~\gev)=0.596$.

\begin{figure}[t]
\begin{center}
\includegraphics[bb=0cm 17cm 30cm 30cm, scale=0.65]{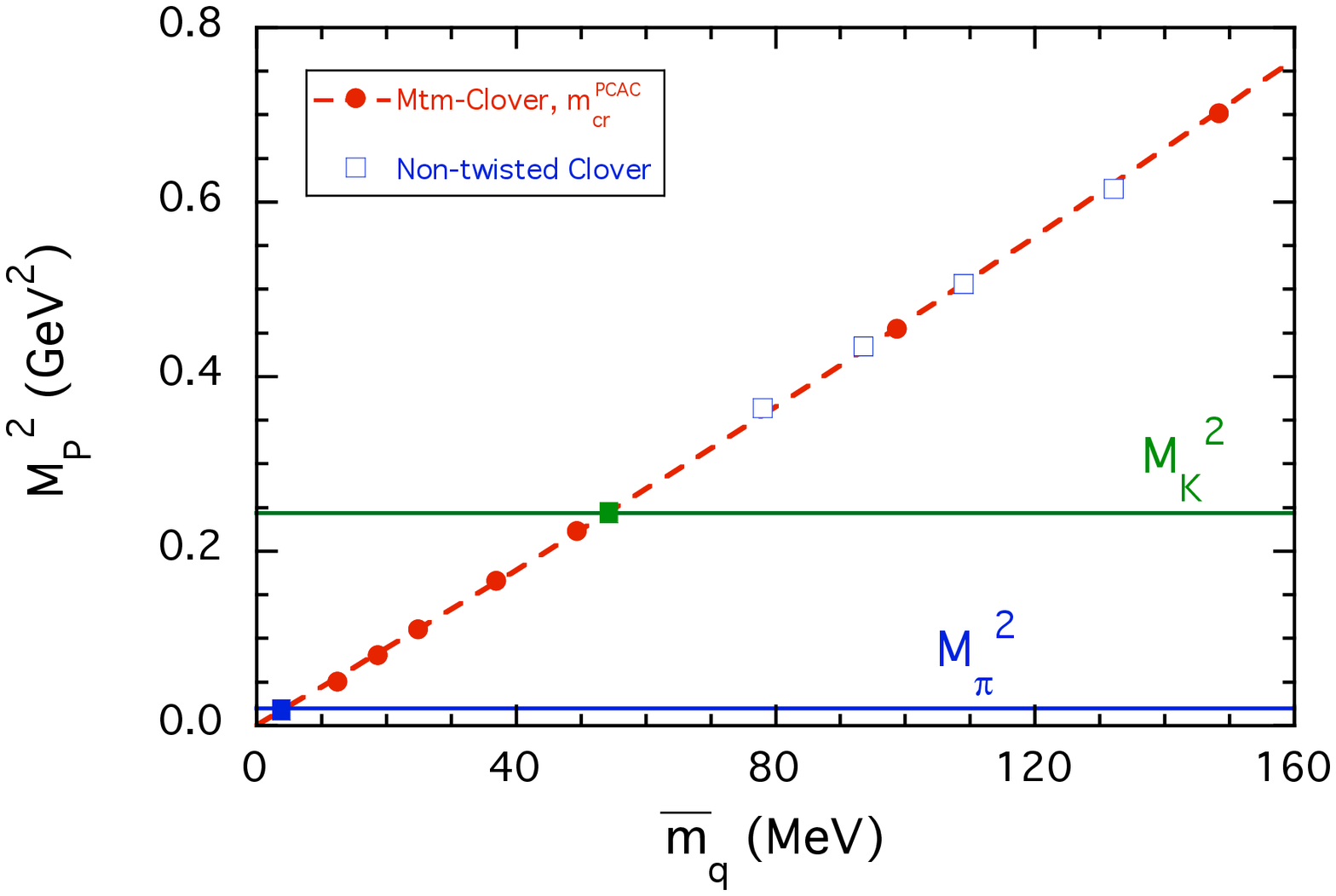}
\end{center}
\caption{\small\sl Pseudoscalar meson masses squared, in physical units, as
a function of the quark mass $\overline{m}_q$ renormalized in the
$\overline{\rm MS}$ scheme at the scale $\mu=2$ GeV. The dashed line
represents the result of the fit to Eq.~(\ref{eq:mpfit}). In the $M_P^2$
values shown in the plot the pure lattice artifact contribution represented
by the coefficient $P_1$ in Eq.~(\ref{eq:mpfit}) has been subtracted. Solid
lines indicate the physical values of the pion and kaon masses. The results
obtained with the non-twisted Clover action in Ref.~\cite{Clover} are also
shown for comparison.}
\label{fig:qmass}
\end{figure}

The physical values of strange and of the average up/down quark masses have
been extracted by fitting the results for the pseudoscalar meson masses
collected in Table~\ref{tab:masses} as a function of the twisted quark
mass. We find that these results are consistent with a quadratic dependence
of the form
\be
(a M_P)^2 = P_1 + P_2 \, (a\mu) +  P_3 \, (a\mu)^2 ~.
\label{eq:mpfit}
\ee
The coefficient $P_1$ parameterizes small, though non-vanishing,
$\mc{O}(a^2)$-discretization effects: we obtain $P_1=\{6(2),5(2),3(2)\}
\cdot 10^{-3}$ in simulations a), b) and c) respectively. In the range of
our simulated quark masses we do not observe any deviation from the simple
quadratic dependence of $M_P^2$ on the quark masses which could signal the
contribution of non analytical chiral logs. In Fig.~\ref{fig:qmass} we show
that the Mtm-LQCD indeed allows a much better control over the extrapolation
to the physical pion mass than it was the case with the standard non-twisted
Wilson/Clover action.

Our final results for the average up/down and strange quark masses are
collected in Table~\ref{tab:quarkmasses}. The determinations obtained
with the various Mtm-LQCD actions are consistent with each other and in
agreement with most of the existing quenched estimates available in
literature (for a recent review see e.g. Ref.~\cite{Sachrajda:2006wp}).
\begin{table}[t]
\vspace{-0.2cm}
\begin{center}
\begin{tabular}{||l||c||c||}
 \hline
 $~~~~~~~~~$Mtm-LQCD          & $m_\ell$ (MeV) & $m_s$ (MeV)\\ \hline
\hline
 Mtm-Clover, $m_{cr}^{\rm PCAC}$ & $4.3~(2)$  & $105~(5)$\\ \hline
 Mtm-Clover, $m_{cr}^{\rm pion}$ & $4.2~(2)$  & $103~(5)$\\ \hline
 Mtm-Wilson, $m_{cr}^{\rm opt}$ & $4.2~(2)$  & $103~(6)$\\ \hline
\hline
\end{tabular}
\end{center}
\vspace{-0.3cm}
\caption{\small\it Values of the average up/down ($m_\ell$) and of the
strange ($m_s$) quark masses renormalized in the $\overline{MS}$ scheme at
the scale of 2 GeV as obtained using the various Mtm-LQCD actions at
$\beta=6.0$.}
\label{tab:quarkmasses}
\end{table}

\subsection{Isospin breaking effects \label{sec:neutral}}
The twisted mass formulation of LQCD breaks both parity and flavour
symmetries at finite lattice spacing. At maximal twist, these breakings
are expected to be of ${\cal O}(a^2)$.

A clear manifestation of isospin breaking effects is a difference between
the neutral and the charged pion mass. In the Mtm-Wilson case, this
splitting has been investigated in Ref.~\cite{pisplit}. The main
observations are: i) isospin breaking effects are numerically large, even
for reasonably fine lattice spacings; quantitatively, it is found $r_0^2\,
(M_{\pi^0}^2 -M_{\pi^+}^2)=c\, (a/r_0)^2$ with $c \approx 10$. ii) The
contribution of disconnected diagrams helps in significantly reducing this
splitting; without adding the disconnected piece, the value of $c$ is about
twice bigger. iii) The optimal choice of the critical mass, which minimizes
parity breaking effects, does not help in reducing discretization effects
related to isospin breaking.

We cannot make the study along the lines presented in Ref.~\cite{pisplit},
since we explore the benefits of using the Mtm-Clover action at single
lattice spacing. Furthermore, we neglect the contribution of disconnected
diagrams, the calculation of which is computationally expensive and
statistically noisy. One observes that the pure connected contribution
provides by itself the neutral-charged mass splitting between pseudoscalar
mesons composed by different, though mass-degenerate, isospin doublets,
namely $P^i = \bar q \frac{\tau^i}{2} \gamma_5 q^\prime$ with $q\neq
q^\prime$. In the absence of isospin breaking effects the neutral and
charged mesons would be degenerate, and thus their splitting still
represents a measure of isospin breaking discretization effect of ${\cal
O}(a^2)$.

\begin{figure}[t]
\begin{center}
\includegraphics[bb=0cm 17cm 30cm 30cm, scale=0.65]{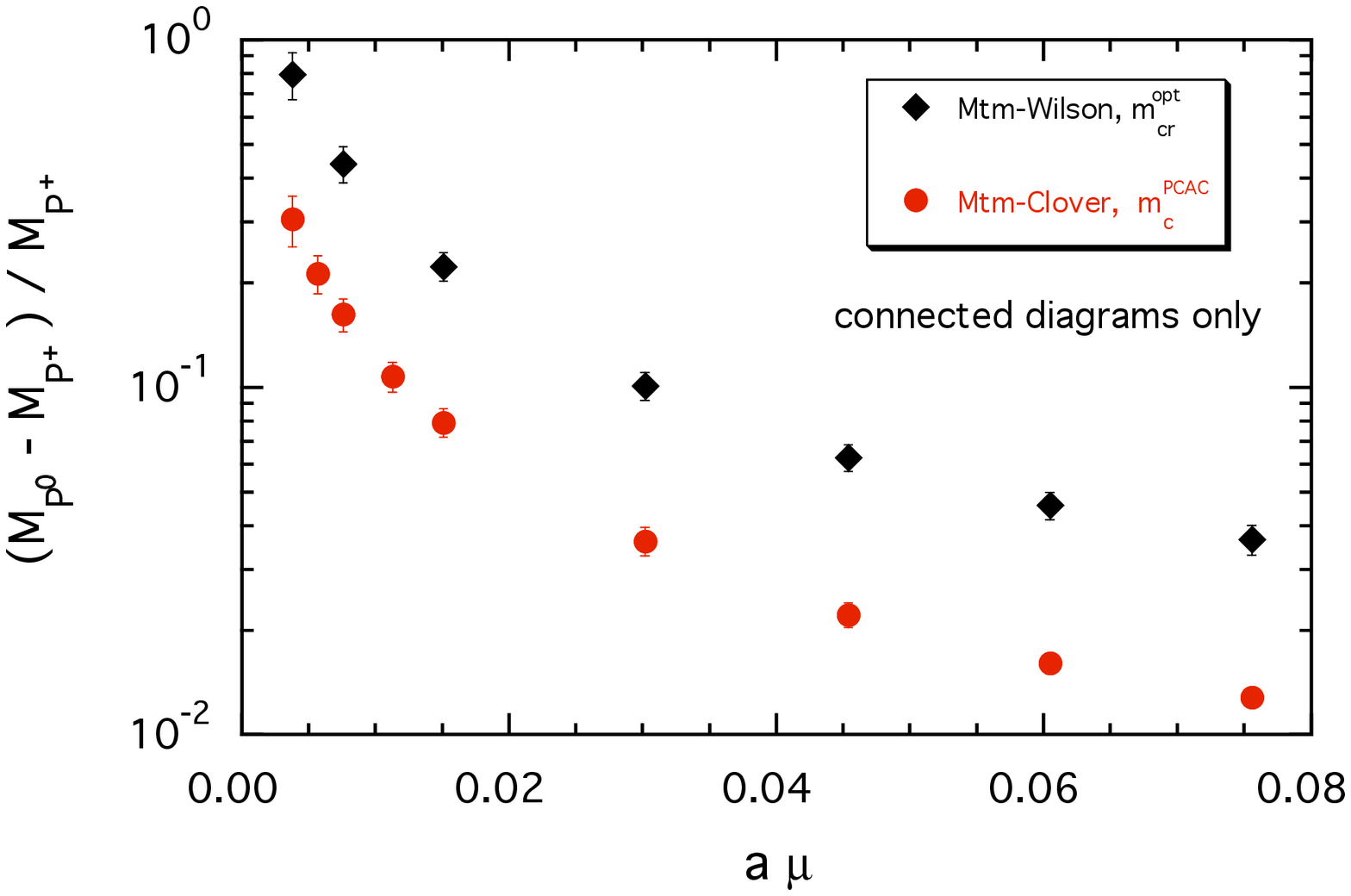}
\end{center}
\caption{\small\it Relative mass splitting between charged and neutral
pseudoscalar mesons as obtained by using the Mtm-Wilson and the Mtm-Clover
actions. The results are presented as a function of the bare quark mass.
For both actions, the calculation of the neutral meson mass only includes
the contribution of connected diagrams (see text for explanations).}
\label{fig:isospin}
\end{figure}

In Fig.~\ref{fig:isospin} we compare the results for the neutral-charged
mass splitting obtained by using the Mtm-Wilson and the Mtm-Clover actions,
with the optimal and the PCAC determination of the critical mass
respectively. In the Mtm-Clover case, the results obtained with the two
different choices for the critical mass ($m_{cr}^{\rm PCAC}$ and
$m_{cr}^{\rm pion}$) do not show significant differences. From
Fig.~\ref{fig:isospin}, it appears that in the Mtm-Clover case the isospin
breaking effects are significantly reduced with respect to the situation in
which the Mtm-Wilson action is used, at least for the pure connected
contribution considered in this study and in the quenched approximation.

\section{Conclusions \label{sec:conclusions}}
In this paper we have presented an exploratory numerical study of Mtm-LQCD
with the Clover term in the action. We have performed quenched simulations
with both Wilson and Clover twisted fermions at $\beta = 6.0$, with
pseudoscalar meson masses ranging from about 1.1 GeV down to approximately
280 MeV. We have investigated, in particular, the role of the Clover term in
reducing the large (IRD) cutoff effects which affect, in general,
simulations with Mtm-LQCD when the quark mass becomes of ${\cal
O}(a\Lambda_{QCD}^2)$. As anticipated in Ref.~\cite{ird}, we find that
adding the Clover term to the twisted mass quark action is very effective in
reducing these effects, as much as using Mtm-Wilson fermions with the
optimal choice of the critical mass.

By using Mtm-Clover fermions, we have also performed other interesting
numerical studies. We have computed the renormalization constant $Z_V$,
finding good agreement with the estimates obtained with the standard Clover
action. We have presented a calculation of the strange and the average
up/down quark masses, which uses non-perturbative renormalization and much
lighter valence quark masses that could not have been simulated with
standard Wilson/Clover fermions, due to the problem of exceptional
configurations. Finally, we have calculated the contribution of connected
diagrams to the charged-neutral pion mass splitting, finding that the use of
the Mtm-Clover action is also beneficial in reducing cutoff effects related
to isospin breaking, at least for this pure connected contribution in the
quenched approximation. We believe that all these results encourage to
pursue this project in exploring the unquenched QCD dynamics at small quark
masses.

\section*{Acknowledgments}
We warmly thank R.~Frezzotti, M.~Papinutto, C.~Pena and G.~Rossi for useful
discussions on the subject of this paper. We also thank the $\XLF$
Collaboration for having provided us with their numerical results obtained
in the simulation with twisted Wilson fermions at maximal twist.

\end{document}